\documentclass[fleqn,usenatbib]{mnras}

\usepackage{newtxtext,newtxmath}
\usepackage[T1]{fontenc}
\usepackage{ae,aecompl}
\usepackage[draft]{todonotes}   % notes showed

\usepackage{graphicx}	% Including figure files
\usepackage{amsmath}	% Advanced maths commands
\usepackage{amssymb}	% Extra maths symbols

\title[dark passengers in stellar surveys]{Dark Passengers\thanks{The dark passenger is a motif in the Dexter book series by Jeff Lindsay. It is describe as a separate entity in the protagonist's psyche that forces him to behave in an abnormal way, and is the result of a sever childhood trauma. We thought the analogy is appropriate for the astrophysical case.} in Stellar Surveys}

\author[Yalinewich et al.]{
Almog Yalinewich,$^{1}$\thanks{E-mail: almog.yalin@gmail.com}
Paz Beniamini,$^{2,3}$
Kenta Hotokezaka$^{4}$
and Wei Zhu$^{1}$
\\
$^{1}$Canadian Institute for Theoretical Astrophysics, 60 St. George St., Toronto, ON M5S 3H8, Canada\\
$^{2}$Department of Physics, The George Washington University, Washington, DC 20052, USA\\
$^{3}$Astronomy, Physics and Statistics Institute of Sciences (APSIS) \\
$^{4}$Department of Astrophysical Sciences, Peyton Hall,
Princeton University, Princeton, NJ 08544, USA
}

\date{Accepted XXX. Received YYY; in original form ZZZ}

\pubyear{2015}

\begin{document}
\label{firstpage}
\pagerange{\pageref{firstpage}--\pageref{lastpage}}
\maketitle

\begin{abstract}
We develop stellar population models to predict the number of binaries with a single luminous member is Gaia and Hipparcos. Our models yield dozens of detections of black hole - luminous companion binaries (BHLC) and hundreds to thousands of neutron star - luminous companion binaries (NSLC) with Gaia. Interestingly, our models also yield a single detection of BHLC binary with Hipparcos, and a few NSLC binaries. We also show how the statistical distribution of detected binaries with a single luminous companion can be used to constrain the formation process of neutron stars and black holes.
\end{abstract}


%%%%%%%%%%%%%%%%%%%%%%%%%%%%%%%%%%%%%%%%%%%%%%%%%%

%%%%%%%%%%%%%%%%% BODY OF PAPER %%%%%%%%%%%%%%%%%%

\section{Introduction}

The Gaia mission \citep{gaia_col_2016} is expected to provide a picture of our galaxy with unprecedented detail. In this paper we consider the prospect of detecting compact objects in binaries according to the astrometric motion of their luminous companions. These include neutron stars (NSLC) and black holes (BHLC). This idea was originally explored in \citet{mashian_loeb_2017} for BHLC binaries. That work estimated $10^5$ such binaries will be detected by Gaia. Later works included the effects of natal kicks \citep{breivik_et_al_2017} and extinction \citep{yamaguchi_kawanaka_et_al_2018}. Each of these considerations reduces the expected number of BHLC binaries orders of magnitude. We note that each of these effects was considered separately. In this work we include extinction and consider different natal kick prescriptions on the BHLC and NSLC systems for the first time.

A point that was not addressed by previous studies is that those same models can be applied to Gaia's predecessor - Hipparcos \citep{1997A&A...323L..49P, eyer_dubath_et_al_2012}. We note that binaries with a single luminous companion have not been reported in the Hipparcos catalogue \citep{1997A&A...323L..53L}, but this might be due to the fact that they were not expected. To demonstrate the capability of Hipparcos, let us consider the following example: the recently discovered BHLC candidate \citep{2018arXiv180602751T} is at a distance of about 2.4 kpc and its apparent magnitude is about 13. A lower limit on the total mass of about 5 $M_{\odot}$, combined with a period of about 83 days, yields a lower limit on the semi major axis of about 0.6 AU. Had it been at a distance 0.6 kpc, its magnitude would have been below 10 and its angular separation would exceed 1 mas, so it would have been detected with Hipparcos. This suggests that Hipparcos' probability of detecting such systems is non-negligible.
%\textcolor{green}{because Hipparcos had the ability to detect such systems. For example, a binary consisting of a $10M_\odot$ black hole and a $10M_\odot$ luminous companion should produce an astrometric signal of 1 mas, if the binary is 1 kpc away and the two components are separated by 2 AU. The brightness of the companion ($V\sim7$) and the astrometric signal are within the reach of Hipparcos.}

Some of the uncertainties in modelling the population of BHLC binaries have to do with the formation channels of black holes. The current understanding is that stellar mass black holes supersede massive stars, as a result of a supernova, failed supernova or direct collapse \citep{heger_fryer_et_al_2003, adams_kochanek_2017}. In general, the mass of the compact remnant is understood to increase with the initial mass of the progenitor and to decrease with the metallicity, but according to some models, there might be regions in the parameter space where this trend is reversed \citep{Belczynski_Bulik_et_al_2010}. Other models suggest that black hole formation only occurs for progenitors whose masses lie within a number of isolated mass ranges \citep{2016arXiv160908411M}.

The situation is further confounded by the observational data, which suggest that the distribution of black holes peaks at around 8 $M_{\odot}$, and is deficient in lower masses \citep{ozel_psaltis_et_al_2010, farr_sravan_et_al_2011, kreidberg_bailyn_et_al_2012}.

It might be possible to obtain information about black hole formation channels by considering neutron stars, as both can form in supernova explosions. Early observations suggested that neutron stars receive natal kicks, and that the velocity is distributed according to a Gaussian with zero mean and a dispersion of about $260\mbox{ km s}^{-1}$ \citep{hobbs_lorimer_et_al_2005}. Newer observations suggest there might be another population of neutron stars, of which the kick velocities follow a double peaked Gaussian distribution but with smaller dispersions of just $77\mbox{ km s}^{-1}$ for 0.42 of the systems and $320\mbox{ km s}^{-1}$ for the rest \citep{verbunt_cator_2017}. Double neutron star binaries exhibit a somewhat similar doubly peaked kick distribution, with roughly two thirds of the systems experiencing kicks lower than $30\mbox{ km s}^{-1}$ and others exhibiting larger kicks, consistent with the regular neutron star population \citep{beniamini_piran_2016, tauris_kramer_et_al_2017}. It is not unreasonable to assume that black holes experience similar natal kicks, although there is no consensus about the magnitude of those kicks. 

To explore these ideas, we developed a Monte Carlo simulation that generates a synthetic population of binaries. The properties of the binaries in the synthetic population were determined according to empirical distribution functions. We then filtered out only those binaries that can be detected by either the Hipparcos or Gaia mission.

This paper is organised as follows. In section \ref{sec:synth_pop} we describe the model according to which we generate a synthetic population of stars. In section \ref{sec:filtering} we describe the conditions according to which whether a given system can be detected by either mission. In section \ref{sec:results} we present the results and we conclude in section \ref{sec:summary}.

\section{Synthetic Population Model} \label{sec:synth_pop}

In this section we describe the different models we use in order to randomise a synthetic population of NSLC and BHLC binaries. The exact details are explained in the original papers \citep{mashian_loeb_2017,breivik_et_al_2017,yamaguchi_kawanaka_et_al_2018}, but we repeat the important relations for completeness.

\subsection{Binary Fraction}
We assume that the binary fraction is 50\% and remains constant regardless of the mass of the primary \citep{2012ASPC..465..284S,yamaguchi_kawanaka_et_al_2018}.

\subsection{Initial Primary Mass}

We assume a Salpeter initial mass function $\frac{d N}{d M} \propto M^{-2.35}$ \citep{salpeter_1955} for the initial mass of the primary. For BHLC we focus on systems of which the primary initial mass is in excess of 20 $M_{\odot}$, as lower mass stars are assumed not to produce black holes. For NSLC we focus on system where the initial mass of the primary is between 8 and 20 $M_{\odot}$. More sophistical variations of this model exist in the literature \citep[e.g.,][]{kroupa_2001, chabrier_2003}, but the main difference between them and the classical Salpeter initial mass function is in the range of low masses, whereas the focus of this work is in the high mass range, where all models have the same slope.

\subsection{Initial mass ratio}

The initial mass ratio between the companion and the primary stars, $q\leq 1$, is assumed to be drawn from a uniform distribution \citep{sana_mink_et_al_2012, duchene_kraus_2013}. The maximum value of the ratio is one when both primary and companion have the same mass, and the minimum is determined from the condition that the companion is more massive than 0.08 $M_{\odot}$, i.e. the dividing line between stars and brown dwarfs \citep{kumar_1962}.

\subsection{Initial Semi Major Axis}

The distribution of the initial semi major axis $A$ is logarithmically flat \citep{sana_mink_et_al_2012, duchene_kraus_2013}. This distribution is sometimes referred to as {\"O}pik's law. The maximum value of the semi major axis is about 1000 AU. Binaries with a larger separation will be torn apart by the tidal forces of the galaxy \citep{connelley_et_al_2008}. The minimum separation is the Roche radius of the primary in the main sequence phase (see section \ref{sec:main_sequence_stellar_radii} for more details).
We note that since this distribution is so wide, the results depend only weakly on the exact value of the cut-off. 

\subsection{Initial Eccentricity}
We assume that all binaries start out with a flat eccentricity distribution in the range $0<e<1$ \citep{duchene_kraus_2013}.

\subsection{Position in the Galaxy}
We only consider stars in the disc component of the galaxy. The distribution of stars follows $\rho_0\exp \left[-\left(r-r_0\right)/h_r-z/h_z\right]$ where $z$ is the distance from the galactic plane, $r$ is the distance from the symmetry axis of the galaxy, $r_0 \approx 8 kpc$ is the distance from Earth to the galactic centre, $\rho_0=0.1$ pc$^{-3}$ is the local stellar number density, $h_z = 250 \, \rm pc$ and $h_r = 3.5 \, \rm kpc$ \citep{bahcall_soniera_1980}. The last coordinate is the rotation angle along the galactic centre, between the line connecting the star to the rotation axis of the galaxy, and the line connecting our solar system to the rotation axis of the galaxy. This angle is assumed to be distributed uniformly between 0 and $2 \pi$.

\subsection{Relic Mass}

To estimate the relic mass we rely the results of numerical simulations of \cite{Belczynski_Kalogera_2008}. Progenitors with masses below 20 $M_{\odot}$ yield neutron stars with a roughly constant masses around 1.5 $M_{\odot}$. Above 20 $M_{\odot}$ a black hole forms instead of a neutron star, and its mass $M_r$ can be related to the mass of the progenitor through \citep{yamaguchi_kawanaka_et_al_2018}
\begin{equation}
M_{r} = 2 \frac{\ln \left(M_1/M_{\odot}-19\right)}{\ln 3} +2\, .
\end{equation}

We note that although most neutron stars have masses that are close to the value we chose, a few have much smaller mass. One example is PSR J1518+4904, which mass is below 1.17 $M_{\odot}$ \citep{janssen__et_al_2008}, and also below the Chandrasekhar mass.

\subsection{Age and Lifetime}

We assume a constant star formation rate throughout the lifetime ($10^{10}$ years) of the galaxy \citep{belczyncksi_kalogera_bulik_2002}. To calculate the main-sequence lifetime of each star we use the model by \citet{hurley_et_al_2000} with metalicity $Z$=0.02. The lifetime of the system as a BHLC binary is therefore approximately the main-sequence lifetime of the secondary minus the main-sequence lifetime of the primary.

\subsection{Main Sequence Stellar Radii} \label{sec:main_sequence_stellar_radii}

The relation between the stellar mass $M$ and stellar radius $R$, for masses in excess of $1.7 M_{\odot}$, is given by \citep{demircan_kahraman_1991}

\begin{equation}
R = 1.6 \left(\frac{M}{M_{\odot}}\right)^{0.83} R_{\odot}
\end{equation}

We use the radius of the companion to check for Roche lobe overflow after the primary collapsed to a black hole. Such a system will either appear as an X ray binary, or will undergo common envelope evolution, which will reduce the period to below the detectable range.
We also use it to filter out binaries whose initial separation is smaller than the main sequence Roche radius of the primary. Binary interaction at this early age and that close will either lead to a merger \citep{sana_mink_et_al_2012} or a surviving binary on an orbit too tight to be detected.

\subsection{Apparent Magnitude and Extinction}

Since the Gaia band is similar to the standard Johnson $V$ band, we will use the latter to calculate the magnitudes. The relation between stellar mass and absolute magnitude ($M_V$) is given by the table in \cite{pecaut_mamajek_2013}. To account for the extinction effect, we assume that the magnitude increases by one for every kpc \citep{trumpler_1930}. In this work we neglect the contribution of post main sequence phases (see detailed discussion in the appendix \ref{sec:post_main_sequence})

\subsection{Binary Evolution}

Binaries with a small enough initial semi major axis exchange mass and alter their orbital parameters. This critical separation depends on the maximum radius the primary attains after the main-sequence stage. As a conservative limit we consider here a lower limit for this radius, we take the radius of the largest known star, UY Scuti, which is 1700 $R_{\odot}$ \citep{arroto_torres_et_al_2013}\footnote{In fact the maximum radius is expected to depend only weakly on the stellar mass, and the limit above, provides a good estimate for the true value}.

In the case of a binary with similar initial individual masses, the mass transfer will start out unstable, but will stabilise when the companion becomes more massive than the primary. We adopt the threshold at a mass ratio $q=0.5$ \citep{yamaguchi_kawanaka_et_al_2018}. The ratio between the final and initial semi major axes is \citep{yamaguchi_kawanaka_et_al_2018}
\begin{equation}
\frac{A_f}{A_i} = \frac{1+2 q + k}{2 k^{3/2} (q+1) \left(\frac{1-k}{2 q}+1\right)^3}
\end{equation}
where $k = M_r/M_1$ is the ratio between the remnant and primary masses. As a result of the mass transfer, the mass of the companion increases by
\begin{equation}
\Delta M_2 = 0.5 \left(1-k\right) M_1
\end{equation}
In the case of a binary with very different masses (i.e. initial mass ratio smaller than 0.5) the mass transfer is unstable, and the companion spirals into the envelope of the primary and expels it. The ratio between the final and initial semi major axes is given by \citep{yamaguchi_kawanaka_et_al_2018}
\begin{equation}
\frac{A_f}{A_i} = \left[\frac{2 \left(1-k\right)}{\alpha \lambda r_l k q} + \frac{1}{k}\right]^{-1}
\end{equation}
where $r_l = R_L/A_i$, $R_L$ is the radius at which Roche lobe overflow occurs and $\alpha \lambda$ is the efficiency of expelling the envelope of the primary \citep{ivanova_justham_et_al_2013}. We consider two variants: $\alpha \lambda = 1$ and $\alpha \lambda = 0.1$, as these values bind most values inferred from observations \citep{de_marco_passy_et_al_2011}.

\subsection{Natal Kicks}

There is little observational information for natal kicks of black holes. There is, however, information about natal kicks for neutron stars. The observational distribution function of the magnitude of the velocity peaks at zero, and declines like a Gaussian with a width of 260 km/s \citep{hobbs_lorimer_et_al_2005}. The orientation is also random, with uniform distribution for the angle. In this work we take after \cite{breivik_et_al_2017, janka_2013} and explore three variations for the kick model
\begin{enumerate}
\item No kick velocity
\item Kick velocity with the same momentum as neutron stars
\item Kick velocity with the same velocity as neutron stars
\end{enumerate}
We assume that prior to the kick, both stars in the binary moved in circular orbits. The effect on the semi major axis and eccentricity are calculated using conservation of angular momentum and energy from the moment after the collapse to the eventual Keplerian orbit \citep{postnov_yungelson_2014}. We further assume that due to the short lifetime of the primary, other changes in the orbital parameters can be neglected.

\section{Detection Criteria} \label{sec:filtering}

In the previous section we described how we generate a synthetic population of binaries. In this section we describe the necessary conditions for detection by either Gaia or Hipparcos.

\subsection{Physical Constraints}

\subsubsection{Age and Lifetime}
The lifetime of the system as a NSLC or BHLC must be larger than its age.

\subsubsection{Eccentricity}
We assume that a binary retains its initial eccentricity unless it enters a common envelope phase, in which case the eccentricity vanishes. Another way in which the eccentricity can change is as a result of natal kicks. Furthermore, if the eccentricity becomes larger than unity, then the two stars are no longer mutually bound.

\subsubsection{Semi Major Axis}
We filter out cases where the terminal semi major axis is smaller than the Roche radius. Closer binaries are assumed to be either X ray binaries, or to have merged. In either case, such binaries are irrelevant for this study. 

\subsection{Hipparcos}

\subsubsection{Magnitude}
Hipparcos can observe stars up to magnitude 12.4.

\subsubsection{Mission Duration and Cadence}
The Hipparcos mission lasted for about 4 years, during which it revisited the observed objects 110 times. Therefore, binaries with periods above about two years and below about a month will not be identified.

\subsubsection{Precision of Proper Motion}
Even when a a luminous companion in a binary is detected, and the orbital parameters determined, the uncertainty in the mass of the dark companion might prevent ruling out compact objects other than a black hole, i.e. neutron stars and white dwarfs.
Given the period $P$, distance to the binary $d$, ratio between semi major axis and distance, $\theta$ and the mass of the companion $M_2$, the mass of the remnant $M_r$ is obtained by solving
\begin{equation}
\left(\frac{P}{1 \rm year}\right)^{-2} \left(\frac{\theta d}{1 \rm AU}\right)^{3} = \frac{M_r^3}{\left(M_r +M_2 \right )^2 M_{\odot}}
 \, .
\end{equation}
The remnant mass $M_r$ depends on four variables: $M_2$, $d$, $\theta$ and $P$. The relative (or logarithmic) uncertainty of the first two scales as $\Delta \theta$ (since the distance depends on parallax and the companion mass depends on distance), while the relative uncertainty of the two others scales as $\Delta \theta/\theta$ as they are pertinent to the binary. Since $\theta$ is a small number, we neglect the uncertainties in $M_2$ and $d$ and only consider the uncertainties in $\theta$ and $P$. In most cases the companion mass is expected to be much larger than the remnant, in which case $M_r \propto \theta/P^{2/3}$ and so the relative uncertainty in remnant mass is given by
\begin{equation}
\frac{\Delta M_r}{M_r} \approx \frac{5}{3}\frac{\Delta \theta}{\theta} \, . \label{eq:dm_vs_dq}
\end{equation}
we adopt a detection threshold of a single standard deviation. For Hipparcos, the angular precision ranges from about 0.7 to 3 mas, depending on the apparent magnitude. We also make the conservative assumption that the critical mass above which a dark companion can only be a black hole is 3 $M_{\odot}$ \citep{yamaguchi_kawanaka_et_al_2018}. Some studies suggest that this mass might be closer to 2 $M_{\odot}$ \citep{lawrence_tervala_et_al_2015, margalit_metzger_2017} and future detection of gravitational waves from merging neutron stars might place more stringent constraints on this value \citep{2018arXiv180109972H}. The criterion for identifying a black hole is therefore $M_r - \Delta M_r > 3 M_{\odot}$.

For NSLC binaries, the criterion for the mass is that it has to be above the Chandrasekhar mass 1.4 $M_{\odot}$ and also below the mass of the most massive neutron star ever detected - PSR B1957+20, with a mass of about $2.4 M_{\odot}$ \citep{kerkwijk_breton_kulkarni_2011}. The criterion for identifying a neutron star is therefore $2.4 M_{\odot} - \Delta M_r > M_r > 1.4 M_{\odot} + \Delta M_r$.

\subsection{Gaia}

\subsubsection{Magnitude}
Gaia can observe stars up to magnitude 20.

\subsubsection{Mission Duration and Cadence} \label{sec:gaia_cadence}
The Gaia mission is planned to last for five years, but can possibly be extended by one to four more years \citep{gaia_col_2016}. In this study we assume a duration of just five years. During this period, it will revisit each patch of sky about seventy times on average. For this reason, like Hipparcos, Gaia will not identify binaries with orbital periods above a few years and below a month.
Our choice for lower cutoff for the period is the same as \citep{yamaguchi_kawanaka_et_al_2018}, but different from \citep{breivik_et_al_2017}, who considered periods as low as half a day. While the Gaia data can be used to detect binaries with periods shorter than the cadence, in such cases the period cannot be estimated accurately. For this reasons we do not consider binaries with a period shorter than the cadence in this study.

\subsubsection{Precision of Proper Motion}
The precision of proper motion of Gaia is between 50 and 100 times better than that of Hipparcos. The actual value of the uncertainty depends on the magnitude of the luminous companion \citep{gaia_col_2016}. We adopt a single standard deviation as a threshold. The relation between precision of proper motion and uncertainty in mass are given in equation \ref{eq:dm_vs_dq}.

\section{Results} \label{sec:results}
In this work we describe a fiducial model similar to the ``curved'' model described in \citet{yamaguchi_kawanaka_et_al_2018}. We prefer this model as our reference because it yields a black hole mass distribution that is similar to observations and is more physically motivated. We explore, in addition, three variations to this model. The first variant is low $\alpha \lambda$, where we use $\alpha \lambda = 0.1$ instead of $\alpha \lambda = 1$ in our fiducial model. The two other variants involve natal kicks, one with the same velocity distribution as neutron stars, and another with the same momentum distribution.

The break-down of the different models and the main results are summarised in the table \ref{tab:results}. All models considered predict hundreds of BHLC binaries and thousands of NSLC binaries with Gaia. The same models predict a marginal detection of a BHLC binary with Hipparcos, and a few NSLC binaries. Our results for BHLC detections with Gaia are in accord with with \citep{yamaguchi_kawanaka_et_al_2018}. In the next subsections we discuss the statistical properties of the different species.

\subsection{Black Holes}

Figure \ref{fig:bh_mass} shows the distribution of detected black hole masses for each of the models. In all models most detected holes have masses between 5 and $8 M_{\odot}$. The reason for that is that on the one hand very massive black holes are rare and on the other, for low mass black holes it is difficult to rule out the possibility that they may be neutron stars.

Figure \ref{fig:companion_mass} shows the distribution of companion masses for different models. We notice two populations in this distribution. The first is massive stars in binaries where common envelope never occurs, and the second is low mass stars where common envelope does occur and the binary becomes very tight. For this reason, reduction of the value of $\alpha \lambda$ completely eliminates the low mass companions with tight orbits. Likewise, kicks are more effective at eliminating binaries with low mass companions.

Figure \ref{fig:periods} shows the distribution of periods for each of the models. In the case of the fiducial model, the distribution of periods is roughly uniform. Reducing the $\alpha \lambda$ parameter eliminates the tight low mass binaries, and natal kicks eliminate some of all binaries, but the low mass tight binaries are more sensitive.

Figure \ref{fig:distances} shows the distribution of  distances from earth. All models peak at about 8 kpc, in accordance with \citet{yamaguchi_kawanaka_et_al_2018}. 

For completeness, we've also included distributions of directly measured quantities in appendix \ref{sec:bhlc_distributions}. These are the magnitudes of the companions in figure \ref{fig:magnitudes}, the ratios between the semi major axes and distance in figure \ref{fig:angles} and the amplitudes of the radial velocities in figure \ref{fig:velocities}.

Our distribution of black hole masses is similar in shape to \citep{breivik_et_al_2017}, but our distribution of companion masses is different. Specifically, their companion mass distribution peaks at low masses and declines with mass, and ours exhibits two populations. The main differences are that we take into account extinction and omit binaries with periods shorter than a month, while \citep{breivik_et_al_2017} considers binaries with periods as short as 0.5 day (see section \ref{sec:gaia_cadence}). This explains why we predict much fewer detections than \citep{breivik_et_al_2017}, and in particular faint low mass stars.

\subsection{Neutron Stars}
Figure \ref{fig:ns_companion_mass} shows the distribution of companion masses for NSLC binaries. Like for the case of black holes this distribution shows two population: massive star which haven't undergone common envelope and post common envelope low mass companions on tight orbits.

Figure \ref{fig:ns_period} shows the period distribution in NSLC binaries. Like black holes, this distribution is relatively flat.

For completeness we've also included in appendix \ref{sec:nslc_distributions} the distribution of distances from earth (figure \ref{fig:ns_distance}), apparent magnitudes of the luminous companions (figure \ref{fig:ns_magnitude}), ratios between the semi major axes and distances (figure \ref{fig:ns_angles}) and radial velocity amplitudes (figure \ref{fig:ns_velocity}).

\begin{table}
\centering
\caption{Parameters of the different models and number of predicted detections of black hole - luminous companion binaries, for both Gaia and Hipparcos missions. \label{tab:results} }
\begin{tabular}{@{}lllll@{}}
                 & fid & al01 & nkm      & nkv      \\ 
$\alpha \lambda$ & 1   & 0.1  & 1        & 1        \\
kick model       & 0   & 0    & Momentum & Velocity \\
Gaia BH          & 150 & 90  & 80      & 50      \\
Hipparcos BH     & 1.6 & 1.6  & 2.1      & 2.8      \\ 
Gaia NS          & 2150 & 1000 & 700 & 500 \\
Hipparcos NS     & 3.4 & 3.8 & 2.4 & 2.1 \\
\end{tabular}
\end{table}

\def\ModelDescription{Blue is the fiducial model, orange is  low $\alpha \lambda$, green is the kick model with the same momentum as neutron stars, and red is kicks with the same velocity.}

\begin{figure}
\includegraphics[width=0.9\columnwidth]{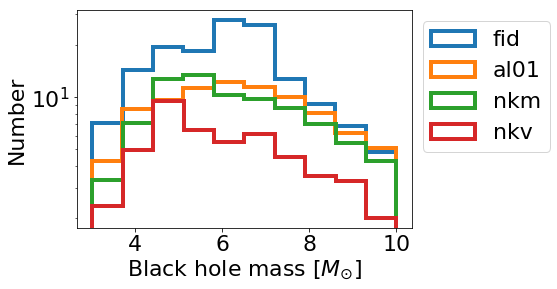}
\caption{
Distribution of masses for black holes observable by Gaia, for each of the models. \ModelDescription
\label{fig:bh_mass}
}
\end{figure}

\begin{figure}
\includegraphics[width=0.9\columnwidth]{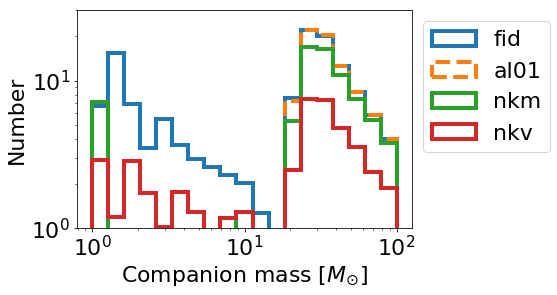}
\caption{
Distribution of companion masses in BHLC binaries detectable with Gaia. \ModelDescription
\label{fig:companion_mass}
}
\end{figure}

\begin{figure}
\includegraphics[width=0.9\columnwidth]{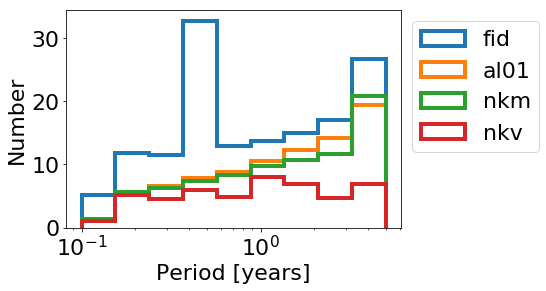}
\caption{
Distribution of periods for BHLC binaries detectable with Gaia. \ModelDescription
\label{fig:periods}
}
\end{figure}

\begin{figure}
\includegraphics[width=0.9\columnwidth]{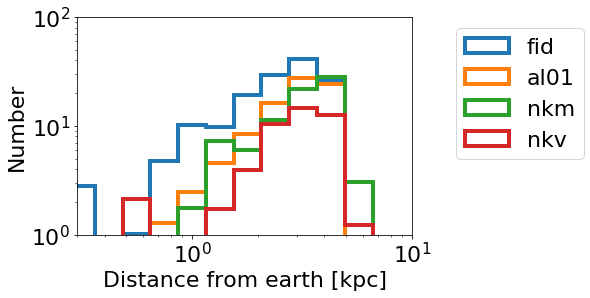}
\caption{
Distribution of distance from earth for BHLC binaries detectable with Gaia. \ModelDescription
\label{fig:distances}
}
\end{figure}

\begin{figure}
\includegraphics[width=0.9\columnwidth]{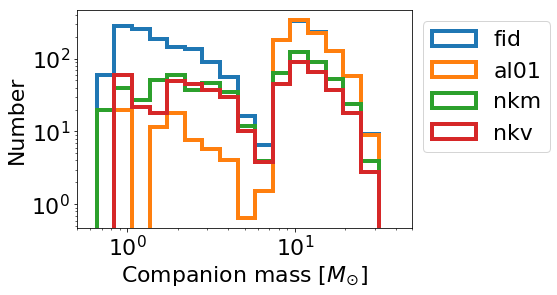}
\caption{
Distribution of companion masses for NSLC binaries detectable by Gaia. \ModelDescription
\label{fig:ns_companion_mass}
}
\end{figure}

\begin{figure}
\includegraphics[width=0.9\columnwidth]{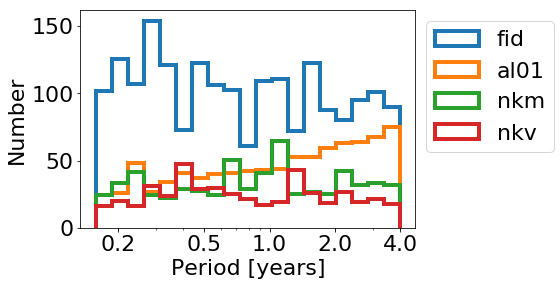}
\caption{
Distribution of periods for NSLC binaries detectable by Gaia. \ModelDescription
\label{fig:ns_period}
}
\end{figure}

\section{Conclusions} \label{sec:summary}
In this work we give some predictions for the detection of binaries with a single luminous companion with Gaia. Similar endeavours exist in the literature, but two novel ideas that we introduce in this paper are (i) that similar methods can be used to detect neutron star - luminous companion systems (which are expected to be more numerous) and (ii) that the data from Hipparcos can be used to constrains the theoretical models.

Our fiducial model predicts a few dozen BHLC detections with Gaia, and about one detections with Hipparcos. The last result is consistent with non detection by Hipparcos, due to uncertainties in the models and statistical fluctuations. All other variants of the fiducial model have similar properties. The corresponding predictions for NSLC binaries are hundreds to thousands with Gaia and a few with Hipparcos. We emphasise that the numbers reported here only apply to binaries where the Chandrasekhar mass is beyond the dark companion's mass error margin.

One of the challenges in this kind of work is estimating the uncertainty. This is because this work is largely based on poorly understood physics and empirical models calibrated with incomplete data. One way to get an estimate for the uncertainty is to explore the sensitivity of the results to changes in the different model parameters. This was done in \citep{yamaguchi_kawanaka_et_al_2018}, and the conclusion was that while the number of BHLC binaries might change by a factor of order unity, the shapes of the distribution functions do not. The results from Gaia, as well as future observations, are necessary to refine these models and improve their precision.

To date, there are no reports of quiescent neutron stars (i.e. neither pulsars nor accreting) in binaries. As for BHLC binaries, there is one reported detection of a $\ge 2.5 M_{\odot}$ mass black hole candidate with a red giant companion \citep{2018arXiv180602751T}. We note that another isolated black hole candidate with mass of about 4 solar masses was detected in the globular cluster NGC 3201 \citep{giesers_dreizler_et_al_2018}. This discovery is less relevant to this study, because stars in a cluster are thought to have formed in a relatively short period of star formation, whereas we are interested in the case where star formation proceeds continuously. Another important difference is that we neglect gravitational interaction between stars that are not mutually bound, which is justified throughout most of the galaxy, but not in dense clusters.

Finally, this study predicts that by the end of its mission, Gaia will provide a great wealth of information on relics of supernova remnants and the stellar collapse. More interestingly, this study also suggests that there could be NSLC and BHLC binaries that can be characterised in the Hipparcos catalogue. To date, most of the $10^5$ Hipparcos objects were identified, but about $10^1$ objects without a valid astrometric solution \citep{1997A&A...323L..49P,martin_mignard_froeschle_1997,1998A&A...330..585M,martin_mignard_et_al_1998} and so the predicted NSLC binaries could be some of those objects.

\section*{Acknowledgements}

AY would like to thank Jo Bovy, Natasha Ivanova, Norm Murray and Dovi Poznanski for the useful discussion, and the anonymous referee for helping improve the paper. WZ would like to thank Cristobal Petrovich for discussions. In this work we used the Matplotlib \citep{hunter2007matplotlib} and NumPy \citep{oliphant2006guide} python packages.

%%%%%%%%%%%%%%%%%%%%%%%%%%%%%%%%%%%%%%%%%%%%%%%%%%



\bibliographystyle{mnras}
\bibliography{references} % if your bibtex file is called example.bib

%\begin{thebibliography}{99}
%\end{thebibliography}

%%%%%%%%%%%%%%%%%%%%%%%%%%%%%%%%%%%%%%%%%%%%%%%%%%

%%%%%%%%%%%%%%%%% APPENDICES %%%%%%%%%%%%%%%%%%%%%

\appendix

\section{Post Main Sequence} \label{sec:post_main_sequence}

Throughout the paper, we assume that the companion is in the main sequence phase. Post main sequence phases like the red giant branch and the asymptotic giant branch are much shorter on the one hand, but much more luminous on the other hand, so their contribution is not immediately obvious.

The probability to detect a star at a certain phase is proportional to the the time it spends in that phase $T$. At distances smaller than the scale height of the galactic disc the probability also scales with luminosity as $L^{3/2}$. However, at such small distances there are not so many stars to boot (about $10^6$) and they detectable with Hipparcos.

At distances larger than the disc scale height, the probability for detection scales as $L T$, or the total energy emitted at a certain phase. For stars more massive than about $2 M_{\odot}$, the radiated energy is dominated by the main sequence phase. Stars less massive than about 0.5 $M_{\odot}$ have a lifetimes longer than the age of the universe.

At such low companion masses, binaries with a large separation that do not undergo common envelope evolution have periods that are too large. Binaries that do undergo common envelope evolution  end up too close so their period is too short. This is evident from the low mass cutoff in figure \ref{fig:companion_mass}.

Another problem with the post main sequence phase is that, due to larger radius of the post main sequence companion, tight binaries may experience Roche lobe overflow. This effect further suppresses the contribution of post main sequence and black hole binaries.

\section{Statistical Distributions of Detected BHLC binaries} \label{sec:bhlc_distributions}

Given below are some statistical distribution of directly measured properties of BHLC binaries according to the different models considered.

\begin{figure}
\includegraphics[width=0.9\columnwidth]{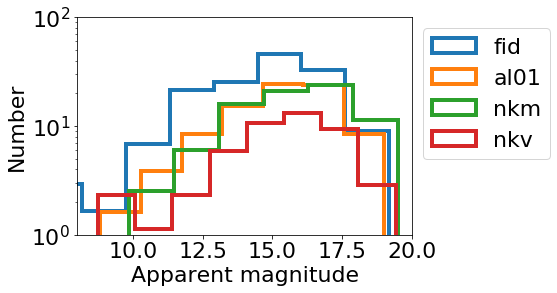}
\caption{
Distribution of apparent magnitudes of the companion in BHLC binaries detectable with Gaia. \ModelDescription
\label{fig:magnitudes}
}
\end{figure}

\begin{figure}
\includegraphics[width=0.9\columnwidth]{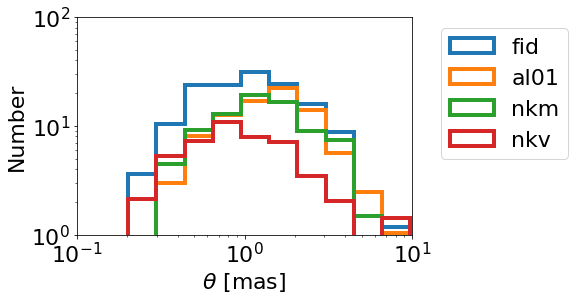}
\caption{
Distribution of ratios between the semi major axes and distances for BHLC binaries detectable with Gaia. \ModelDescription
\label{fig:angles}
}
\end{figure}

\begin{figure}
\includegraphics[width=0.9\columnwidth]{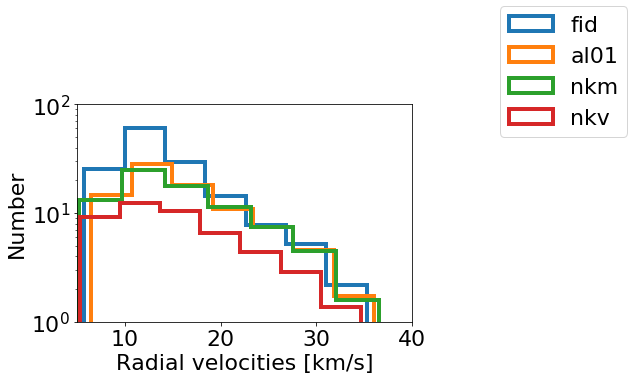}
\caption{
Distribution of radial velocity amplitudes for BHLC binaries detectable with Gaia. \ModelDescription
\label{fig:velocities}
}
\end{figure}

\section{Statistical Distributions of Detected NSLC binaries} \label{sec:nslc_distributions}
Given below are some statistical distribution of directly measured properties of NSLC binaries according to the different models considered.

\begin{figure}
\includegraphics[width=0.9\columnwidth]{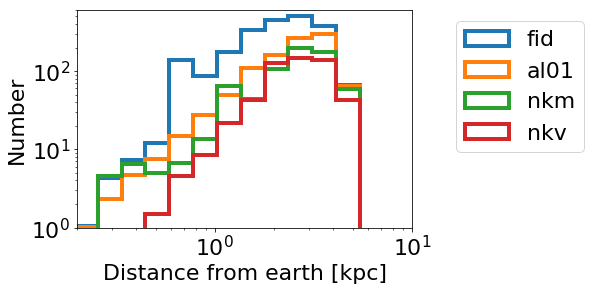}
\caption{
Distribution of distances from earth for NSLC binaries detectable by Gaia. \ModelDescription
\label{fig:ns_distance}
}
\end{figure}

\begin{figure}
\includegraphics[width=0.9\columnwidth]{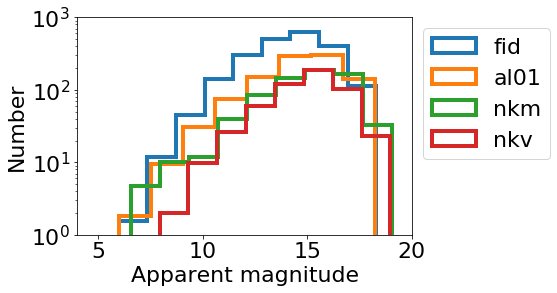}
\caption{
Distribution of apparent magnitudes for NSLC binaries observable by Gaia, for each of the models. Blue is the fiducial model, orange is low $\alpha \lambda$, green is the kick model with the same momentum as neutron stars, and red is kicks with the same velocity.
\label{fig:ns_magnitude}
}
\end{figure}

\begin{figure}
\includegraphics[width=0.9\columnwidth]{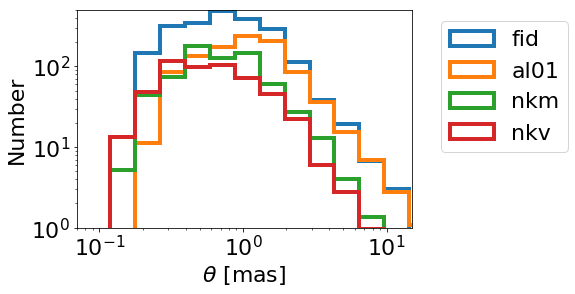}
\caption{
Distribution of ratios between the semi major axes and distances for NSLC binaries observable by Gaia, for each of the models. Blue is the fiducial model, orange is low $\alpha \lambda$, green is the kick model with the same momentum as neutron stars, and red is kicks with the same velocity.
\label{fig:ns_angles}
}
\end{figure}

\begin{figure}
\includegraphics[width=0.9\columnwidth]{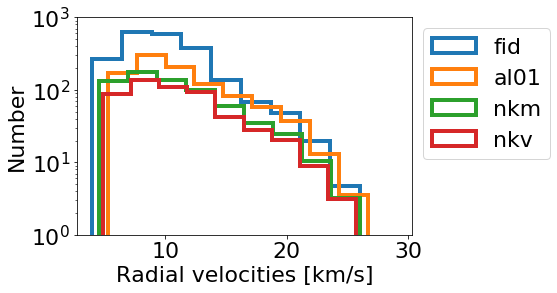}
\caption{
Distribution of radial velocity amplitudes for NSLC binaries observable by Gaia, for each of the models. Blue is the fiducial model, orange is low $\alpha \lambda$, green is the kick model with the same momentum as neutron stars, and red is kicks with the same velocity.
\label{fig:ns_velocity}
}
\end{figure}

%%%%%%%%%%%%%%%%%%%%%%%%%%%%%%%%%%%%%%%%%%%%%%%%%%

% Don't change these lines
\bsp	% typesetting comment
\label{lastpage}
\end{document}